\documentclass[fleqn,usenatbib]{mnras}

\usepackage{newtxtext,newtxmath}
\usepackage[T1]{fontenc}

\DeclareRobustCommand{\VAN}[3]{#2}
\let\VANthebibliography\thebibliography
\def\thebibliography{\DeclareRobustCommand{\VAN}[3]{##3}\VANthebibliography}

\usepackage{graphicx}	% Including figure files
\usepackage{amsmath}	% Advanced maths commands
\usepackage{xcolor}
\usepackage{afterpage}
\usepackage{tikz}
\usetikzlibrary{positioning}

\newcommand\gta{\lower 0.5ex\hbox{$\buildrel > \over \sim\ $}} %greater than about
\newcommand\lta{\lower 0.5ex\hbox{$\buildrel < \over \sim\ $}} %less than about

\newcommand{\Pwd} {P_\mathrm{WD}}
\newcommand{\Pms} {P_\mathrm{MS}}
\newcommand{\Pclass} {P_\mathrm{class}}
\newcommand{\Fbeta} {F_\beta}
\newcommand{\Gbp}{G_\mathrm{BP}}
\newcommand{\Grp}{G_\mathrm{RP}}

\title[Data-Driven Classification of WD Stars]{Data-Driven Selection and Spectral Classification of White Dwarf Stars}

\author[Vincent et al.]{
Olivier Vincent\thanks{E-mail: o.vincent@umontreal.ca}, 
P. Bergeron %$^{1}
and P. Dufour%$^{2,3}$
\\
D\'epartement de Physique, Universit\'e de Montr\'eal, C.P.~6128, Succ.~Centre-Ville, Montr\'eal, Qu\'ebec H3C 3J7, Canada\\
}

\date{Accepted XXX. Received YYY; in original form ZZZ}

\pubyear{2022}

\begin{document}
\label{firstpage}
\pagerange{\pageref{firstpage}--\pageref{lastpage}}
\maketitle

\begin{abstract}
% set scene
The next generation of spectroscopic surveys is expected to provide spectra for hundreds of thousands of white dwarf (WD) candidates in the upcoming years.
% identify problem
Currently, spectroscopic classification of white dwarfs is mostly done by visual inspection, requiring substantial amounts of expert attention.
% outline how problem is solved + main results
We propose a data-driven pipeline for fast, automatic selection and spectroscopic classification of WD candidates, trained using spectroscopically confirmed objects with available Gaia astrometry, photometry, and Sloan Digital Sky Survey (SDSS) spectra with signal-to-noise ratios $\geq9$.
The pipeline selects WD candidates with improved accuracy and completeness over existing algorithms, classifies their primary spectroscopic type with $\gtrsim 90\%$ accuracy, and spectroscopically detects main sequence companions with similar performance. 
We apply our pipeline to the Gaia Data Release 3 cross-matched with the SDSS Data Release 17 (DR17), identifying 424 096 high-confidence WD candidates and providing the first catalogue of automated and quantifiable classification for 36 523 WD spectra. Both the catalogue and pipeline are made available online.
% link to scene
Such a tool will prove particularly useful for the undergoing SDSS-V survey, allowing for rapid classification of thousands of spectra at every data release.
\end{abstract}

\begin{keywords}
White Dwarfs -- Methods: Data Analysis -- Catalogues -- Surveys
\end{keywords}

%%%%%%%%%%%%%%%%%%%%%%%%%%%%%%%%%%%%%%%%%%%%%%%%%%

%%%%%%%%%%%%%%%%% BODY OF PAPER %%%%%%%%%%%%%%%%%%
%%%%%%%%%%%%%%%%%%%%%%%%%%%%%%%%%%%%%%%%%%%%%%%%%%%%%%
%%%%%%%%%%%%%%%%%%%%%%%%%%%%%%%%%%%%%%%%%%%%%%%%%%%%%%
\section{Introduction}\label{sec:intro}
%Must allow the reader to place the work in context and find key background papers if they want to know more

% Present problem in wide context
The field of astronomy is entering an era of Big Data, with the volume of astronomical data doubling every 16 months and is predicted to keep doing so for the next few years \citep{smith2022}. Classical methods that rely on human supervision and specialist expertise are rapidly becoming insufficient to handle this stream of opportunities, and concerns are arising that they may strongly delay the important discoveries, or worse, completely miss them. Machine learning methods have emerged as a natural response to these concerns and have already become commonplace in many physical sciences \citep[see][for recent reviews]{carleo2019,smith2022}. Stellar science has also recently seen interesting applications of machine learning, including main sequence star spectral classification \citep{sharma2020}, stellar parameter inference \citep{ting2019, chandra2020}, and trigonometric parallax calibration \citep{leung2019}.

% Narrow focus to the specific subtopic of your paper
Until now, the quantity of astronomical data available for the study of white dwarfs has remained small enough to be manageable using classical methods. Most of the spectroscopic data come from SDSS \citep{gunn2006}, which gradually provided optical spectroscopy along with broadband photometry for the majority of the $\sim$33 000 currently known white dwarfs \citep{harris2003, kleinman2004, eisenstein2006, kleinman2013, kepler2015, kepler2016, fusillo2021}. Newer surveys dramatically contrast with this figure, such as the Gaia survey \citep{prusti2016} that measured astrometry for over a billion objects, resulting not only in increasing the number of white dwarfs with parallax measurements by about 3 orders of magnitude \citep[][and references therein]{bedard2017}, but also in the identification of $\sim$260 000 high-confidence white dwarf candidates \citep[][]{fusillo2021}. Available data sources also worth mentioning include the Survey Telescope And Rapid Response System photometry \citep[Pan-STARRS;][]{chambers2016}, the Two Micron Sky Survey near-infrared photometry \citep[2MASS;][]{skrutskie2006}, and the spectroscopic data from the Large Sky Area Multi-Object Fiber Spectroscopic Telescope \citep[LAMOST;][]{cui2012}.

% Point out a problem/lack in the literature 
The next generation of observatories and surveys \citep[e.g. SDSS-V, DESI, 4MOST;][]{kollmeier2017,desi2016,dejong2014} will be providing spectroscopic data for a large number of white dwarf candidates, unlocking both unprecedented statistical analyses and detailed studies of white dwarfs. However, extracting the white dwarf observations from the billions of expected spectra will be an impossible task without the aid of automated tools. While machine learning methods for the selection of white dwarf candidates as well as the classification of DA versus non-DA have started to surface \citep{smart2021,sanjuan2022}, the most recent spectroscopic catalogues are still built using visual inspection \citep{fusillo2021, kepler2021}. More generally, studies of large white dwarf samples usually involve a number of manual steps that require substantial amounts of time that experts could and should be spending on more important aspects of the data analysis \citep{caron2022}.

% Define the exact problem and present what you did
As a first step towards addressing this lack of tools, we present a neural network-based pipeline for rapid and automated selection and spectroscopic classification for white dwarf candidates. The pipeline is comprised of three independent modules that identify white dwarf candidates based on Gaia astrometry and photometry, classify the main spectroscopic signature of white dwarf spectra, and detect the presence of spectroscopic contamination by a main sequence star. The entire pipeline is trained using human-labelled spectroscopically confirmed objects, and spectroscopic modules are trained and tested using SDSS Data Release 16 spectra \citep[DR16;][]{ahumada2020}. In the future, each module of the pipeline will be able to serve as a base classification model for various tasks and surveys. As a proof-of-concept, we produce a catalogue containing 1.3M Gaia objects cross-matched with the SDSS Data Release 17 \citep[DR17;][]{abdurrouf2022}, the latest and last data release of SDSS-IV.

% sujet divisé
This paper is structured as follows. In Section \ref{sec:meth}, we describe the different components of the pipeline, as well as the methods and data used to train them. The performance of the pipeline on test data is presented in Section \ref{sec:res}. We also test the spectroscopic modules in different data quality regimes, and on white dwarf spectra with multiple spectral signatures. We then showcase the pipeline by creating a catalogue using 1.3M white dwarf candidates from the Gaia survey cross-matched with the SDSS DR17 in Section \ref{sec:dr17}. Finally, we summarize our work and provide concluding remarks in Section \ref{sec:conc}.

%%%%%%%%%%%%%%%%%%%%%%%%%%%%%%%%%%%%%%%%%%%%%%%%%%%%%%
%%%%%%%%%%%%%%%%%%%%%%%%%%%%%%%%%%%%%%%%%%%%%%%%%%%%%%
\section{Methodology}\label{sec:meth}
\subsection{Pipeline description}

% make large fig
\begin{figure*}
    \centering
    \begin{tikzpicture}[
    % Define nodes
    roundnode/.style={circle, draw=blue!60, fill=blue!5, very thick, minimum size=2cm},
    roundnode2/.style={circle, draw=black!60, fill=white!5, very thick, minimum size=2cm},
    squarednode/.style={rectangle, draw=blue!60, fill=blue!5, very thick, minimum size=1cm},
    bsquarednode/.style={rectangle, draw=blue!60, fill=blue!5, very thick, minimum size=1.5cm},
    bsquarednode2/.style={rectangle, draw=black!60, fill=white!5, very thick, minimum size=1.5cm},
    ]
    %Add nodes
    % node   (label)  [position]  {text inside}
    \node[roundnode2, align=center]     (gaia)                       {Gaia\\parameters};
    \node[roundnode2, align=center]        (spec)      [below=.75cm of gaia]  {Spectrum};
    \node[bsquarednode, align=center]      (cnd)       [right=1.5cm of gaia]  {Module 1:\\Candidate Selection};
    \node[bsquarednode, align=center]     (main)      [right=2cm of cnd]  {Module 2:\\Main Spec. Type};
    \node[bsquarednode2, align=center]      (out)        [right=2cm of main]   {Spectroscopically\\classified WD};
    \node[bsquarednode, align=center]      (ms)        [below=1.25cm of out]   {Module 3:\\WD$+$MS Detection};

    %\coordinate[below=2cm of main] (under_main_mid);
    
    % Draw box
    %\node[draw=red,fit=(image) (kappa4)] (box) {};
    
    % Add text
    \node[text width=3cm] at (7.25,.25) {$\Pwd$};
    \node[text width=3cm] at (11.35,.25) {$\Pclass$};
    \node[text width=3cm] at (13.9,-1.4) {$\Pms$};
    \node[text width=3cm] at (10.4,-2.3) {DA, DB, DC};
    
    %Add lines
    \draw[-, thick, dotted] ([shift={(.25,0.)}]main.south) -- ([shift={(0.25,-1.75cm)}]main.south);
    \draw[->, thick, dotted] ([shift={(.25,-1.75)}]main.south) -- ([shift={(0.,0.25)}]ms.west);

    \draw[->, thick] ([shift={(-0.25,-2.)}]main.south) -- ([shift={(-0.25,0)}]main.south);
    
    \draw[->, thick] (gaia.east) to (cnd.west);
    \draw[->, thick, dotted] (cnd.east) to (main.west);
    \draw[->, thick, dotted] (main.east) to (out.west);
    \draw[->, thick, dotted] (ms.north) to (out.south);
    \draw[->, thick] (spec.east) to (ms.west);
 
    \end{tikzpicture}
    \caption{Schema of the pipeline. For a given object, the Gaia parameters are sent to the candidate selection module to determine the probability of being a white dwarf ($\Pwd$). If this probability meets certain criteria, the spectrum of the object is sent to the first spectroscopic classification module, where the probability that its primary spectroscopic type belongs to one of the 13 possible classes ($\Pclass$) is calculated. If the most probable class is either a DA, DB or DC, the spectrum is also sent to the WD$+$MS module to determine the probability of a main sequence companion ($\Pms$).}
    \label{fig:pipeline}
\end{figure*}
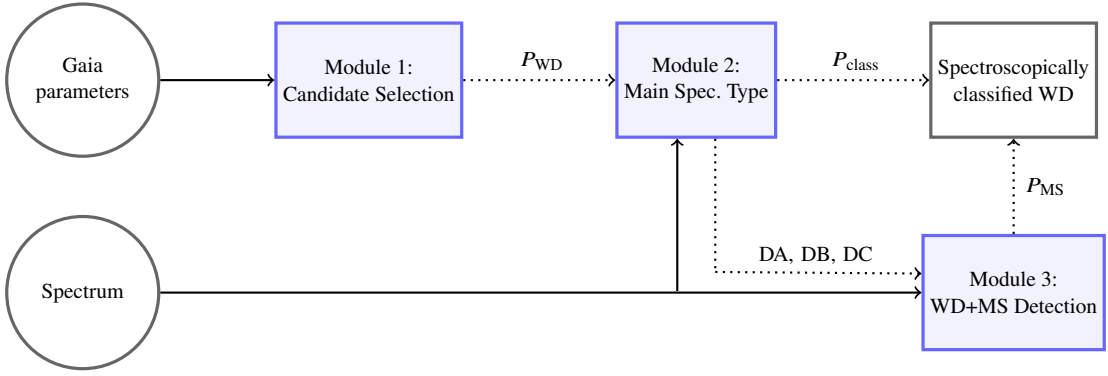

The white dwarf selection and spectral classification pipeline consists of three different modules: one for white dwarf candidate selection among Gaia objects, one for classification of the primary spectroscopic type\footnote{In this paper, we follow the terminology described in \citet{sion1983} where the upper case letter following the D (for degenerate) indicates the primary spectroscopic type in the optical spectrum, and where the following upper case letters indicate the secondary spectroscopic features.}, and one for the detection of spectral contamination from a main sequence companion. The pipeline is illustrated in Figure \ref{fig:pipeline}. Each module consists of 10 neural networks grouped together to form a deep ensemble, which are known to offer improved generalization and performance over single neural networks \citep{lee2015, lakshminarayanan2016}. Since neural networks typically have millions of parameters, there exists a large number of different parameter combinations that might sufficiently approximate the function the networks are trying to model. By ensembling them, we create a distribution of diverse functions from which we can compute statistics for our classifications \citep{fort2019}. For each pipeline module, we take the mean and standard deviation of the predictions of their respective 10 networks as the final prediction and uncertainty, respectively. Within each module, the neural networks are built using an unique architecture (see Appendix \ref{app:arch} for more details), but are trained using different initialization and portions of the training data sets.

The candidate selection module is an ensemble of binary classifiers that take the three Gaia magnitudes ($G$, $\Gbp$, $\Grp$), their flux errors, both proper motion components ($\mu_\alpha$ and $\mu_\delta$), and the parallax measurement, along with all their respective uncertainties. Our choice of input data is based on the white dwarf candidate Random Forest classifier by \citet{smart2021}, and differences between our classifiers are explored in Section \ref{sec:dr17}. Additionally, we include \texttt{phot\_bp\_rp\_excess\_factor} as an input parameter, as we found it helpful for identifying binaries composed of a white dwarf and a main sequence star (WD$+$MS). We also tested the corrected excess factor proposed by \citet{riello2021} and found no notable differences in performance when compared to the uncorrected factor. In total, we use 13 Gaia input parameters to predict the probability $\Pwd$ of an object being a white dwarf.

Spectral classification by the two other modules requires a spectral coverage of at least 3840 to 7000 \AA\ for the determination of the primary spectroscopic type, which we increase to 9000 \AA\ for the detection of contamination from a main sequence companion. The first spectroscopic module outputs the probability $\Pclass$ that the object belongs to one of the 13 following classes: DA, DB, DC, DO, DQ, hotDQ, DZ, DAH, PG1159, cataclysmic variable (CV), sdB, sdO, or sdBO. The white dwarf classes follow the classification system proposed by \citet{sion1983} and \citet{fusillo2021}, whereas subdwarfs follow the system proposed by \citet{geier2017}. The neural networks of this module are trained assuming a multiclass problem, meaning the classes are assumed to be mutually exclusive, and the sum of probabilities must equal unity. We emphasize that this approach does not attempt to classify secondary spectroscopic features, but does, however, provide a consistent primary type for hybrid white dwarfs (see Section \ref{sec:hybrid}). The main sequence companion module, similar to the candidate selection module, is made of binary classifier networks that output a probability $\Pms$ that the spectrum is contaminated by a MS star.

\subsection{Data selection and processing}
To train and test our networks, we made use of all objects with a confirmed spectral type in the Montreal White Dwarf Database \citep[MWDD;][]{dufour2017}, in the Gaia-SDSS catalogue of \citet[][henceforth GF21]{fusillo2021}, as well as the subdwarf star catalogues of \citet{geier2017} and \citet{geier2020}. We downloaded all SDSS DR16 spectra and Gaia DR3 data associated with these objects from the Science Archive Server\footnote{https://www.sdss.org/} and Gaia Archive\footnote{https://gea.esac.esa.int/archive/}. In order to minimize human error in the spectral classifications, we discarded spectra with a signal-to-noise ratio (SNR) lower than 9 between 4500 and 5500 \AA, which are mostly DA and DB white dwarfs, as well as subdwarfs. Each module of the pipeline has further restrictions that are described in Section \ref{sec:res}.

All SDSS spectra go through the following preprocessing. We remove the sky emission lines around 5577  6300, and 6363 \AA\ by replacing a region of 9 pixels centered on those lines with interpolated values of the nearest neighbouring pixel. We then pseudo-continuum normalize the spectra using a running window of 50 pixels width and selecting pixels in the 85th percentile or higher, on which we fit a fourth order Chebyshev polynomial. The pseudo-continuum pixels are restricted to the 3842-7000\AA range for the main spectroscopic normalization module and to 3842-9000\AA for the main sequence companion detection module. Telluric pixels are also excluded.

As a final preprocessing step for the spectroscopic module inputs, we compute the average and standard deviations for all pixels of all continuum-normalized spectra within their respective training sets, and zero-center the spectra by subtracting the average, and then by dividing the standard deviation. The same preprocessing procedure is applied to the Gaia parameters of the candidate selection module by using the mean and standard deviations of each parameter computed over the entire training set.

\section{Pipeline training and validation}\label{sec:res}
%In this section we demonstrate the performance of each component of our classification pipeline on stars with both SDSS and Gaia observations.  %move this to intro?

\subsection{Module 1: Identification of white dwarf candidates}
We begin by looking at the candidate selection module, which takes 13 Gaia parameters as input, and outputs a probability $\Pwd$ for the object to be a white dwarf (see Section \ref{sec:meth}). Our sample includes all objects in the MWDD as well as the GF21 catalogue with available spectral classification. These objects are split into two categories: white dwarfs and non-white dwarfs; this last category also includes subdwarfs and main sequence stars. We apply the following cuts: $G_{\rm abs} > 6 + 5(G_{\rm BP}-G_{\rm RP})$ to remove most main sequence stars, and $\textsc{parallax\_over\_error} > 10^{-3}$ to remove any object with excessive parallax measurement error.

Any object with missing Gaia parameters or any known MS$+$WD binary is also removed, leaving a total of 35 930 stars among which 33 416 are spectroscopically confirmed white dwarfs, and 2514 are confirmed non-white dwarfs. We train the neural networks using a different random selection of 25 000 objects for each of them, validate each one using a different random selection of 5500 objects, and test them with their respective 5430 remaining objects.

\begin{figure}
    \centering
    \includegraphics[width=\columnwidth]{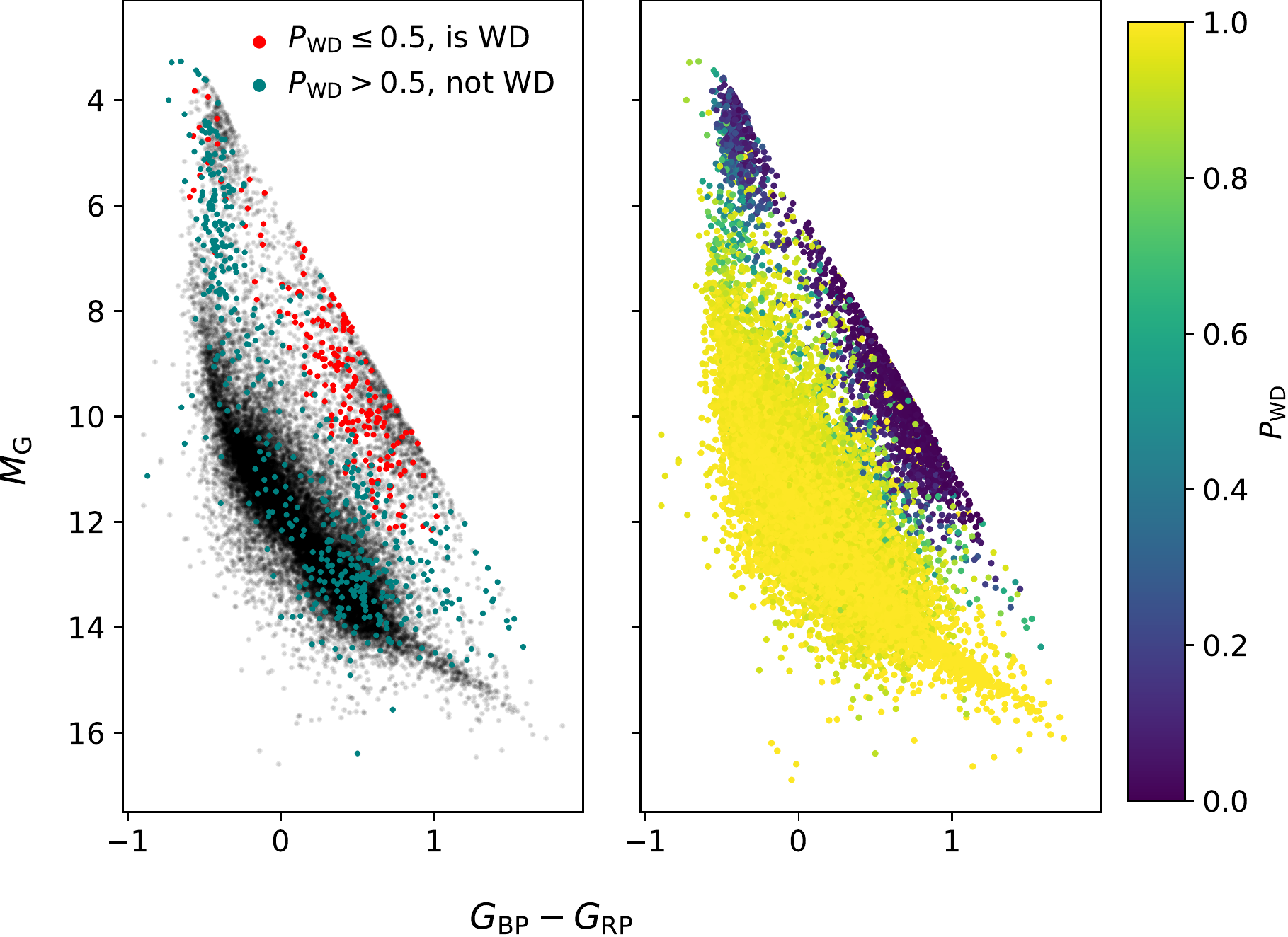}
    \caption{Gaia HRD of the sample used to test the candidate selection ensemble. Left: Red dots represent confirmed white dwarfs for which the ensemble predicts a low $\Pwd$, while teal dots are non-WD objects predicted to have a high $\Pwd$. Objects with correct classifications are colored in grey. Right: Distribution of $\Pwd$ over the entire HRD.}
    \label{fig:hrdwd}
\end{figure}

Using a probability threshold of 0.5, the networks correctly identify, on average, $99.2\%$ of white dwarfs ($\Pwd>0.5$) and $83.5\%$ of non-white dwarfs ($\Pwd\leq0.5$) in their respective test sets. This translates into about $1.4\%$ contamination in objects classified as white dwarfs and 0.9\% white dwarfs being missed by the networks. In order to minimize biases learned by individual networks, we assemble them and use the average probability as the final prediction. We run the entire sample through the ensemble and show the resulting distribution of $\Pwd$ over the Hertzsprung-Russell diagram (HRD) in the right panel of Figure \ref{fig:hrdwd}. Objects located within the white dwarf locus show high probabilities of being a white dwarf, rapidly dropping as we move towards the main sequence.

We also show the HRD of misclassified objects in the left panel of Figure \ref{fig:hrdwd} when applying the $\Pwd=0.5$ threshold, highlighting that most misclassified white dwarfs reside within the edge of the main sequence locus. While it is unsurprising that the ensemble shows confusion in such regions, we note that these misclassified white dwarfs are mostly of hot spectral types (DOA, PG1159) or have photometric effective temperatures above $\gtrsim20 000$\;K \citep{fusillo2021,dufour2017}. Consequently, the ensemble may be less sensitive to very hot white dwarfs found outside the white dwarf locus.

\begin{figure}
    \centering
    \includegraphics[width=\columnwidth]{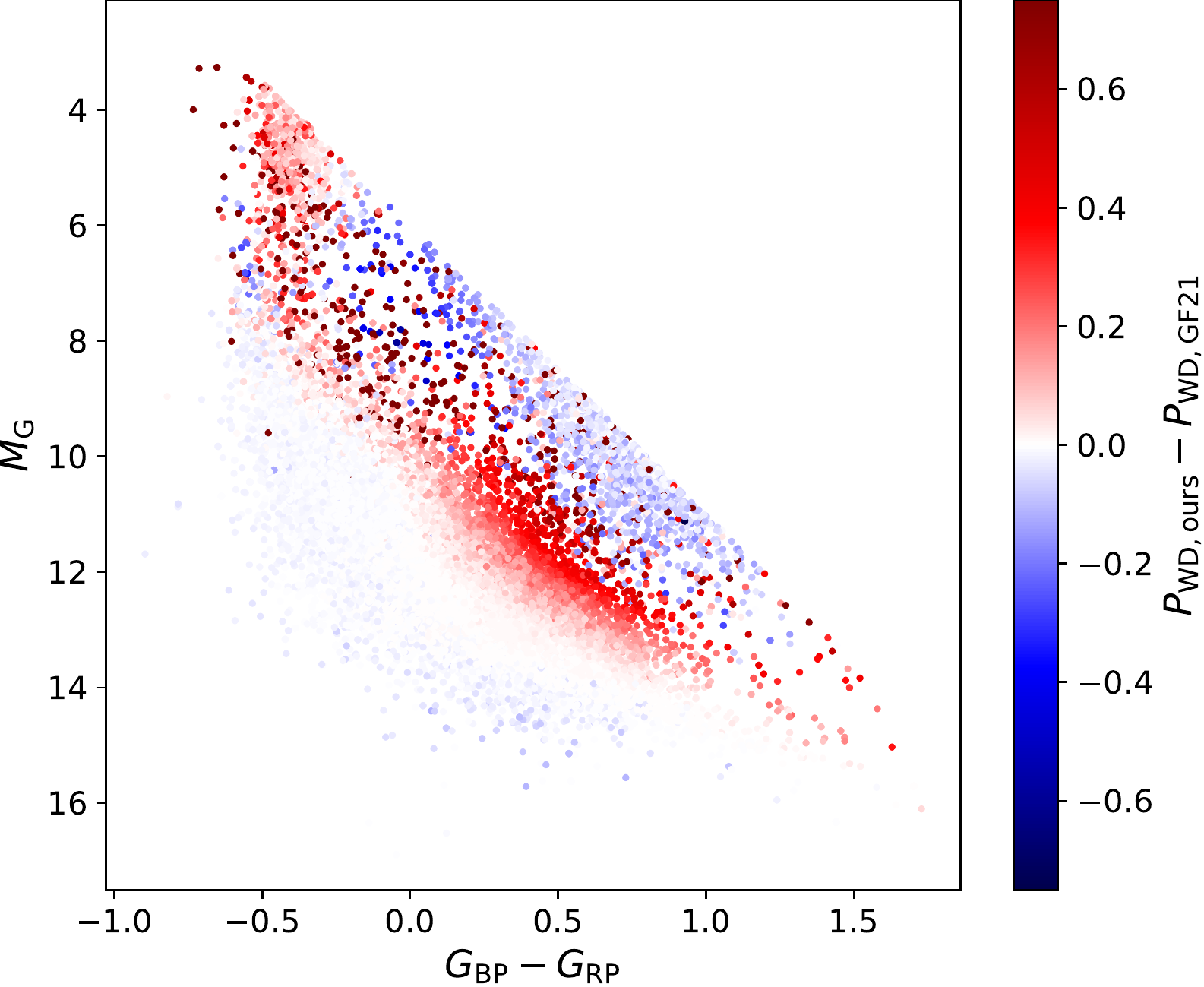}
    \caption{Comparison of the predicted $\Pwd$ by our ensemble and those of \citet{fusillo2021} over the HRD. Red points indicate that the network predicts higher probabilities than GF21, while blue points indicate the opposite.}
    \label{fig:hrdgf}
\end{figure}

We compare the ensemble probabilities with those of GF21 in Figure \ref{fig:hrdgf} by plotting the differences between our $\Pwd$ and theirs over the HRD. We find large differences (over 0.5) between the probabilities for 532 objects, primarily located at the edges of the white dwarf locus and main sequence tail. 471 of these objects are spectroscopically confirmed white dwarfs for which our ensemble predicts high probabilities ($\Pwd\gtrsim0.85$), while the GF21 catalogue indicates very low probabilities ($\Pwd\lesssim0.15$). The remaining 61 objects have subdwarf spectroscopic classifications and show a mix of very high or low $\Pwd$ for both our predictions and those of GF21. Our ensemble shows excellent performance and appears more robust for objects located in ambiguous regions of the HRD. We attribute this to the fact that the neural networks can learn highly non-linear features in a larger parameter space to make their predictions, rather than providing a simple density estimation of the HRD.

\subsection{Module 2: Primary spectroscopic type}

\begin{table}
 \caption{Sample of spectra used to train the networks and summary of the ensemble predictions}
 \label{tab:training}
 \centering
 \begin{tabular}{lrrrr}
    \hline
    Label & $N$ & $N_\mathrm{agree}$ & $N_\mathrm{uncert}$ & $N_\mathrm{disagree}$ \\
    \hline
    DA & 17485  & 17281 & 93 & 111 \\
    DC & 1718   & 1548 & 65 & 105 \\
    DB & 1649   & 1625 & 7 & 17 \\
    DZ & 439    & 424 & 5 & 10 \\
    DQ/DQpec & 296    & 260 & 14 & 22 \\
    DAH & 179   & 137 & 27 & 15 \\
    DO/DAO & 142& 104 & 12 & 26 \\
    hotDQ & 73  & 70 & 2 & 1 \\
    PG1159 & 23 & 15 & 3 & 5 \\
    sdB & 759   & 617 & 51 & 91 \\
    sdOB & 389  & 118 & 35 & 102 \\
    sdO & 255   & 203 & 67 & 119 \\
    CV & 221    & 208 & 5 & 8 \\
    \hline
    Total & 23628 & 22610 & 386 & 632\\
  \hline
 \end{tabular}
\end{table}

This section focuses on the primary spectroscopic type classification module of our pipeline. We train the networks using spectra of non-hybrid objects from the GF21 Gaia-SDSS catalogue and subdwarf catalogues from \citep{geier2017,geier2020}. The number of spectra for each class is listed in Table \ref{tab:training}. In order to increase the size of the DO and DQ classes, we include DAO white dwarfs in the DO class, and DQpec white dwarfs in the DQ class. We do not include objects with spectral classification found only in the MWDD as their types may have been assigned based on observations obtained by other means than the SDSS. We split the 23 628 spectra in Table \ref{tab:training} into 21 265 for the training set, and 2363 for the validation set, ensuring the proportion of each spectral type remains the same. We do not use a testing set due to the very small number of PG1159, hotDQ, and DO stars, and instead cross-validate the networks by using a completely different validation set for each network, i.e.~a different 10\% of the dataset for each one. 

In order to determine whether a spectral classification is to be considered reliable or flagged as uncertain, requiring a visual inspection, we rely on a prediction probability threshold based on the generalized $F_\beta$ score:

\begin{equation}
    F_\beta = (1+\beta^2)\frac{\mathrm{Precision}\times\mathrm{Recall}}{(\beta^2\times\mathrm{Precision}) + \mathrm{Recall}}\;,
\end{equation}

\noindent
where

\begin{equation}
    \mathrm{Precision} = \mathrm{TP / (TP + FP)}\;,
\end{equation}
\begin{equation}
    \mathrm{Recall} = \mathrm{TP / (TP + FN)}\;.
\end{equation}

\begin{figure}
    \centering
    \includegraphics[width=\columnwidth]{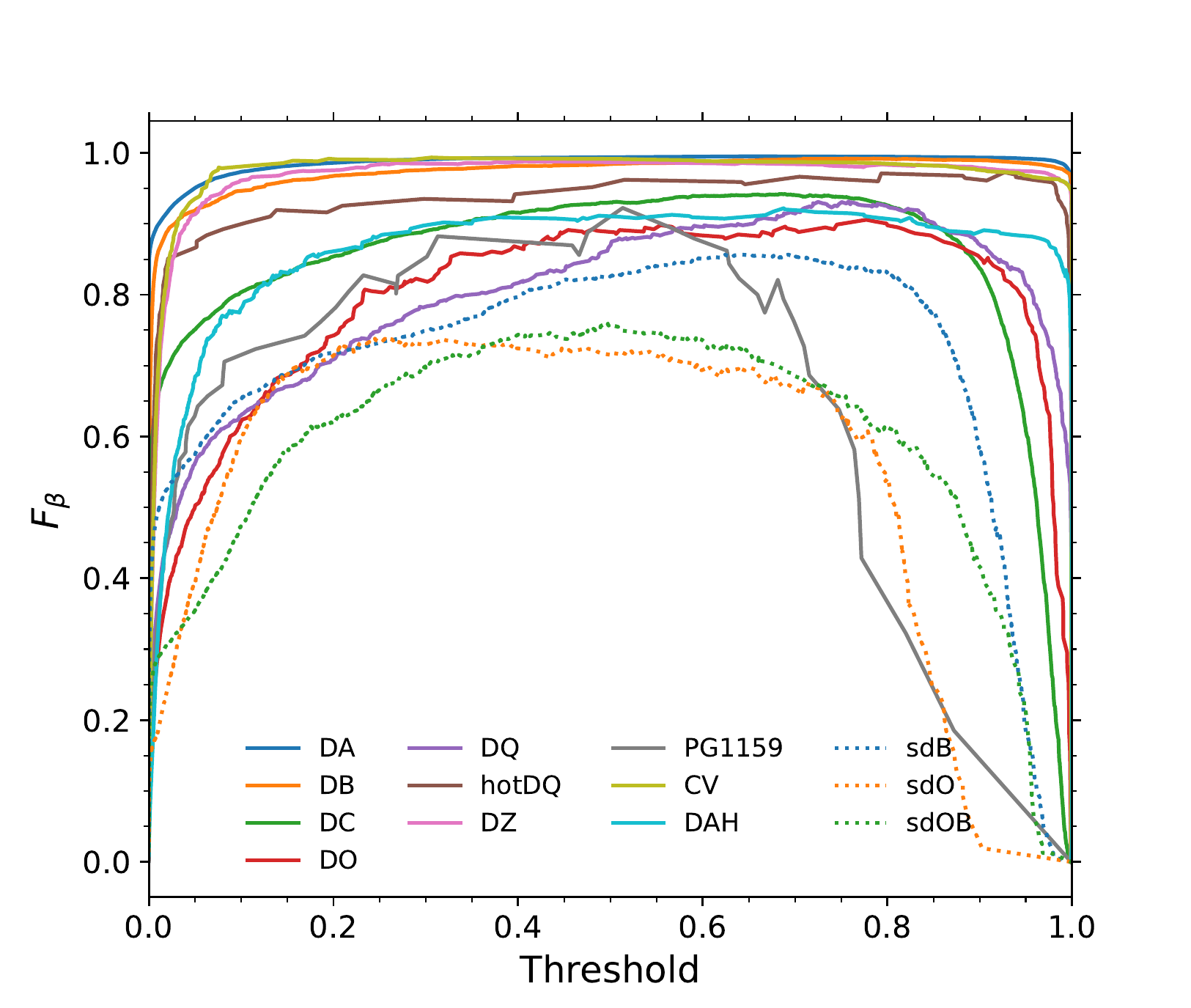}
    \caption{$F_\beta$ score as a function of classification threshold for each class label. The $\beta$ factor is set to 0.5, making precision twice as important as recall (see text).}
    \label{fig:fbeta}
\end{figure}

\noindent Precision is a measure of purity for a given class, while recall is a measure of completeness, and both are calculated using the count of true positives (TP), false positives (FP), or false negatives (FN). Their importance can be weighted in the $F_\beta$ score using the $\beta$ parameter, where $\beta>1$ places more weight on recall, and $\beta<1$ places more importance on precision. As we aim to minimize the need for visual inspection of spectra, we consider precision twice as important as recall and set $\beta=0.5$. We calculate $F_\beta$ for each class for every individual network using their respective validation set, and plot the average $F_\beta$ curve for each class in Figure \ref{fig:fbeta}. We find that a threshold of 0.6 provides the highest score for most classes. Depending on the case studied, different thresholds may prove optimal. For example, if the subdwarf classification is not of interest, a threshold of 0.5 may be more appropriate, or if DQ white dwarfs are the only objects of interest, a threshold of 0.75 may be more useful. In light of these results, we consider a spectrum to be classified if the highest prediction probability for a class is above the threshold $\Pclass\geq0.6$. Objects with no class probabilities above this threshold are tagged for visual inspection. These are discussed at the end of this section.

\begin{figure}
    \centering
    \includegraphics[width=\columnwidth]{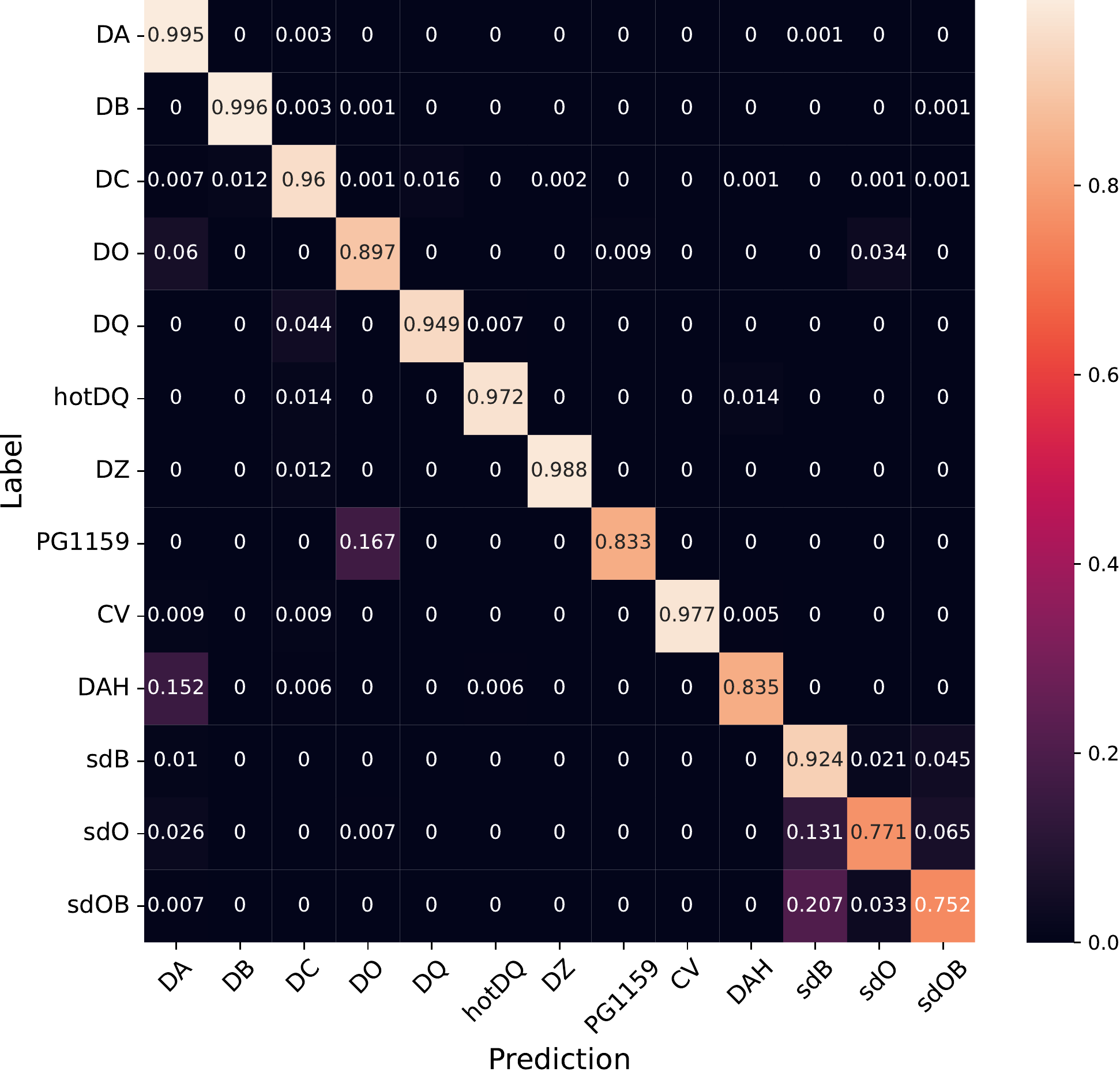}
    \caption{Average confusion matrix of the primary spectroscopic type confidently predicted ($\Pclass\geq0.6$) by the networks for objects with a single known spectroscopic signature. The values are normalized row-wise.}
    \label{fig:cmatrix}
\end{figure}

Having defined what constitutes a confirmed classification, we now perform a cross-validation to verify the performance of the networks for objects with a non-hybrid spectral type (i.e., DA, DB, DQ, DZ, etc.). We calculate the confusion matrix for each of the ten networks with their respective validation set, normalize them row-wise and take their average, producing the confusion matrix shown in Figure \ref{fig:cmatrix}. The majority of white dwarf classes show $\gtrsim$90\% agreement between network predictions and human labels, while PG1159 and DAH show the lowest agreement at $\sim$83\%. The networks predict a PG1159 class for $\sim$13\% of human-labelled DO, and $\sim$17\% of the human-labelled DAH as being DA, lowering the score of their respective types. Subdwarfs appear to be the most confused classes as many sdO and sdBO are predicted to be sdB. Even so, all three classes show good agreement with human labels. In what follows, we ensemble the ten neural networks, predict classes for the entire spectroscopic sample, and review spectra belonging to classes displaying the highest percentage of disagreeing predictions and labels. We list the number of disagreements for each class in Table \ref{tab:training} as $N_\mathrm{disagree}$, along with the number of agreeing predictions/labels ($N_\mathrm{agree}$) and number of spectra flagged for visual inspection ($N_\mathrm{uncert}$).

%DAH
Visual inspection of 15 objects classified as DAH by GF21 but with different predictions reveals that the ensemble tends to confuse DAH with low SNR DA, especially if the blue part of the spectrum is of low quality. In low SNR spectra, magnetic line splitting looks very similar to wide, but noisy, Balmer lines. Such confusion has been shown to be frequent in the study of magnetic white dwarfs by Hardy et al. (in prep.), who have found 400 out of 651 white dwarf spectra to have erroneously been classified as magnetic in previous literature. In our case, however,  human labels appear to be correct, and the ensemble indicates a 5-30\% probability of the objects being a DAH, which can be used as an indicator of weakly detectable magnetic splitting. There is one notable object with differing classifications, Gaia DR3 2849930668862492544, predicted to be a hotDQ, which was classified as a DQA in the in-depth analysis of DQ/DZ white dwarfs by \citet{coutu2019}.

% why DO-PG are confused
A similar investigation of the 3 objects classified as PG1159 by GF21, but with different predictions, points to the presence of ionized helium in the spectra of these objects, a feature that the ensemble strongly associates with the DO class. We emphasize the very small number of known PG1159 stars, and the fact that neural networks are data-driven algorithms that strongly depend on the number of available examples to learn from \citep{he2009,zhu2015}. It is thus not surprising to see a lower performance for the class with the least number of spectra. Furthermore, PG1159 and DO are suspected to share a common spectral evolution pathway \citep{bedard2022}, and so the decision boundary between the two is ill-defined by nature. Objects found on the fine line between DO and PG1159 can usually be identified with high prediction probabilities for both classes, typically $\gtrsim0.6$ for the first, and $\gtrsim0.2$ for the second.

%why sdb are confused
As for the apparent confusion between subdwarf types, we strongly suspect the culprit to be common features among the various types. Indeed, according to the classification scheme proposed by \citet{geier2017}, both sdOB and sdO subdwarfs may show a mix of hydrogen, neutral helium, and ionized helium lines. The considerable feature overlap, along with their occasional presence, of these classes could easily induce confusion to both our ensemble and a human classifier, resulting in errors in not only the predictions but labels as well. While a subtype classification has been proposed by \citet{geier2020} to separate hydrogen-rich from helium-rich spectra, the number of objects for each class would be too small for the networks to learn meaningful features.

% uncertains
The results discussed so far have been restricted to high-confidence predictions, i.e.~those with a class probability above the threshold $P_\mathrm{class}>0.6$. Among the 23 628 spectra used to train and validate the networks, 2.7\% have their highest predictions falling below this threshold. We list the number of uncertain spectra per label in Table \ref{tab:training}. The majority of cases are subdwarfs, consistent with the fact that their classes are the most confusing to the ensemble. Uncertain white dwarf classifications can be grouped into three broad categories: (1) their spectra is close to an ill-defined boundary between multiple classes, (2) the spectra have low SNR in regions where important class-specific features are found, and (3) the spectra may possess an unusual feature among its class. Categories (1) and (2) are self-explanatory and the most affected classes share the same explanations as the misclassifications discussed above. Spectra belonging to category (3) include rare objects such as Gaia DR33985469616188225152, a DQ with oxygen lines \citep{gansicke2010}, or Gaia DR3 3731667388643923840, also a DQ but with metal traces \citep{coutu2019, farihi2022}. A simple approach to filter out the majority of uninteresting spectra with uncertain classifications would be to apply a SNR cut, keeping most spectra in categories (1) and (3).
%A more extensive discussion of unusual objects identified by our ensemble is presented along with our Gaia-SDSS DR17 catalogue in Section \ref{sec:dr17}.

\begin{table}
\caption{Sample of white dwarfs with known secondary spectroscopic features}
\label{tab:subs}
\centering
\begin{tabular}{lrrrr}
    \hline
    Label & $N$ & $N_\mathrm{PIA}$ & $N_\mathrm{uncert}$ & $N_\mathrm{disagree}$ \\
    \hline
    DAB/DBA & 326 & 271 & 20 & 35 \\
    DAZ/DZA & 105 & 92 & 10 & 3 \\
    DBZ/DZB & 97 & 86 & 11 & 0 \\
    DBAZ & 59 & 59 & 0 & 0 \\
    DABZ & 20 & 17 & 3 & 0 \\
    DZBA & 14 & 13 & 1 & 0 \\
    DZAB & 5 & 4 & 1 & 0 \\
    DOZ & 2 & 1 & 1 & 0 \\
    DQZ & 1 & 0 & 1 & 0 \\
    DA$+$MS & 783 & 750 & 22 & 11 \\
    DB$+$MS & 37 & 32 & 1 & 4 \\
    DC$+$MS & 23 & 16 & 4 & 3 \\
    \hline
    Total & 1472 & 1341 & 75 & 56 \\
    \hline
\end{tabular}
\end{table}

\subsection{White dwarfs with secondary spectroscopic features}\label{sec:hybrid}
% wd with subtypes
We now turn to white dwarfs with secondary spectroscopic features, such as DAZ, DBA, etc. Since the networks are trained on objects with only primary spectroscopic types, it is important to asses their predictions when confronted with ambiguous data. Our white dwarf sample with known secondary spectroscopic features is listed in Table \ref{tab:subs} and totals 1472 spectra, most of which are labelled DBA/DAB and WD$+$MS binaries. The second column of Table \ref{tab:subs} lists the number of spectra with a predicted primary spectroscopic type matching one of the human labels of their respective class (Prediction In Any class; $N_\mathrm{PIA}$), while the third and fourth columns list the number of spectra flagged for visual inspection and disagreeing predictions/labels, respectively. The ensemble predictions show excellent agreement with human labels, with $\sim$91\% of all spectra matching one of the possible types and only $\sim$3\% disagreement. Upon visual inspection, nearly all spectra with predictions inconsistent with their labels were found to be misclassified in the GF21 catalogue.

% wd with subtype "misclassifications"
Out of 38 spectra labelled as having secondary spectroscopic features for which the ensemble predicts no matching primary type, only two appear to have been erroneously predicted by the networks. The ensemble prediction for the spectra of Gaia DR3 3781616827503753088 and Gaia DR3 1028779636740111872 points to a DC type instead of either a DA or DB that would match their proposed DBA label, probably due to the weak strength of the lines. As a matter of fact, Gaia DR3 1028779636740111872 was classified as a DB+DC binary by \citet{kleinman2013}, and the helium lines are most likely diluted by the DC companion. 11 other spectra are labelled as DBA/DAB, although they clearly show ionized helium lines, and the ensemble predicts a DO primary class for these spectra, which we confirm to be correct by cross-checking the predictions with the spectral types in the MWDD. 22 spectra are labelled as DBA/DAB/DAZ but are predicted to be subdwarfs by the ensemble, which we also confirm using the MWDD. One spectrum of Gaia DR3 1884548739436672640 shows carbon in its spectrum and is correctly predicted to be a DQ, although it is labelled as a DBA. The spectrum for Gaia DR3 1587462866571138048 is labelled as DZA, while the ensemble predicted a DB type, consistent with the DBZA spectral type in the MWDD. The predictions made by the ensemble for the last two objects are also consistent with the spectral types DQA and DBAZ given by \citet{coutu2019}. The ensemble seems to provide a primary spectroscopic type reliably when classifying white dwarfs with secondary spectroscopic features.

% wd with MS subtype
The situation is quite similar when classifying the 1472 spectra labelled as having a main sequence companion, as most spectra with predictions inconsistent with their labels were found to have wrong labels. We note that although there are differences between the predicted and labelled primary spectroscopic types for the cases discussed below, a main sequence companion is indeed always present. Out of the 18 spectra with disagreeing predictions and labels, the only spectrum with a possible erroneous ensemble prediction belongs to the object Gaia DR3 2464385576553809792, which is predicted to be a DZ but labelled as a DC$+$MS. The spectrum is dominated by the MS companion, showing unusual features to the ensemble, which likely confuse it. Prediction uncertainties for this spectrum are very high, with the DZ type being predicted with 61\% probability, but with 37\% uncertainty. Interestingly, the DC type is predicted with 24\% probability, but with 34\% uncertainty. Such uncertainties may be used to discern spurious high-confidence classifications, and even give a hint as to which class is the correct one. 
9 of 18 spectra labelled as DA$+$MS, but predicted to be CV, all show emission in at least one of their hydrogen lines. Two spectra for the object Gaia DR3 922604914151538816 labelled as DB$+$MS show ionized helium lines and are predicted to be DO, consistent with the DO$+$MS classification in the MWDD. The spectrum for Gaia DR3 1219552974402511104 is labelled as DA$+$MS but is predicted to be a subdwarf, probably due to its narrow hydrogen lines. A spectrum for Gaia DR3 2836746940329352448 labelled as DC$+$MS is predicted to be a DB, consistent with the MWDD, and one available spectrum for Gaia DR3 733784442283662464 labelled DB$+$MS does not show any obvious features, consistent with the ensemble prediction suggesting a DC. Finally, a spectrum for Gaia DR3 2536561952205972608 labelled as a DA$+$MS but predicted to be a DC also does not seem to show any obvious features. However, higher SNR spectra of the same object reveal it is indeed a DA$+$MS. These verifications lend support to the ensemble providing primary spectroscopic types that are more robust than visual inspection for WD$+$MS spectra.

\subsection{Module 3: Main sequence companionship}\label{sec:ms}
The third and final module of the pipeline makes use of SDSS spectra to predict the probability of contamination by a main sequence companion. Since the redder part of white dwarf spectra often contains the most obvious signature of the presence of a main sequence companion, we rely on a wider wavelength coverage than the main classification module, extending to 9000 \AA. We use the GF21 catalogue to form two groups for our training sample: all white dwarfs labelled as having a main sequence companion form the positive group, while those with a DA, DB or DC type form the negative group, totalling 832 and 20 486 spectra for each group, respectively. 

We then train each neural network using the same approach as for the candidate selection module by randomly selecting, for each network, 16 000 and 2700 spectra for training and validation, respectively, keeping the remaining 2618 spectra for testing. Using a probability threshold of 0.5, the networks correctly identify, on average, 91.6\% of WD$+$MS spectra and 99.8\% of uncontaminated white dwarfs. About 0.4\% of spectra labelled as single white dwarfs are predicted to have a companion, and about 4.9\% of spectra labelled as having a companion are predicted to be single white dwarfs. We visually inspected the spectra with disagreeing predictions and labels, and found that 36 out of 76 spectra labelled as having no companion turn out to obviously have one, consistent with the very high probabilities ($\gtrsim$0.8) given by the network. Out of 41 spectra predicted as having no companion but labelled as having one, 9 show obvious companion contamination. Among these, 4 are DB$+$MS and 1 is a DC$+$MS. We find no obvious clues as to why these objects were erroneously classified, but we suspect the type of white dwarf may be the cause, as DB and DC binaries are much less numerous and may be harder for the ensemble to recognize. We also notice 11 CVs and 5 subdwarfs that were mislabelled as WD$+$MS in the GF21 catalogue, for which the networks give very small probabilities of being a WD$+$MS, consistent with their real classification. We find the rest of the spectra with disagreeing predictions and labels too visually ambiguous to determine companionship by visual inspection alone.

\subsection{Performance vs SNR}
The SNR of a spectrum is perhaps one of the most important factors affecting the visibility of spectroscopic features, as well as the overall spectral shape, making the same object look radically different when observed at low and high SNR. Moreover, large surveys generally do not provide uniform SNR distributions of spectra, and so classification algorithms may struggle to classify correctly spectra of objects that have much higher (or lower) SNR than the bulk of spectra of its class. Below, we demonstrate that the two spectroscopic classification modules of our pipeline perform well over the entire SNR range of the SDSS spectra.

\begin{figure}
    \centering
    \includegraphics[width=\columnwidth]{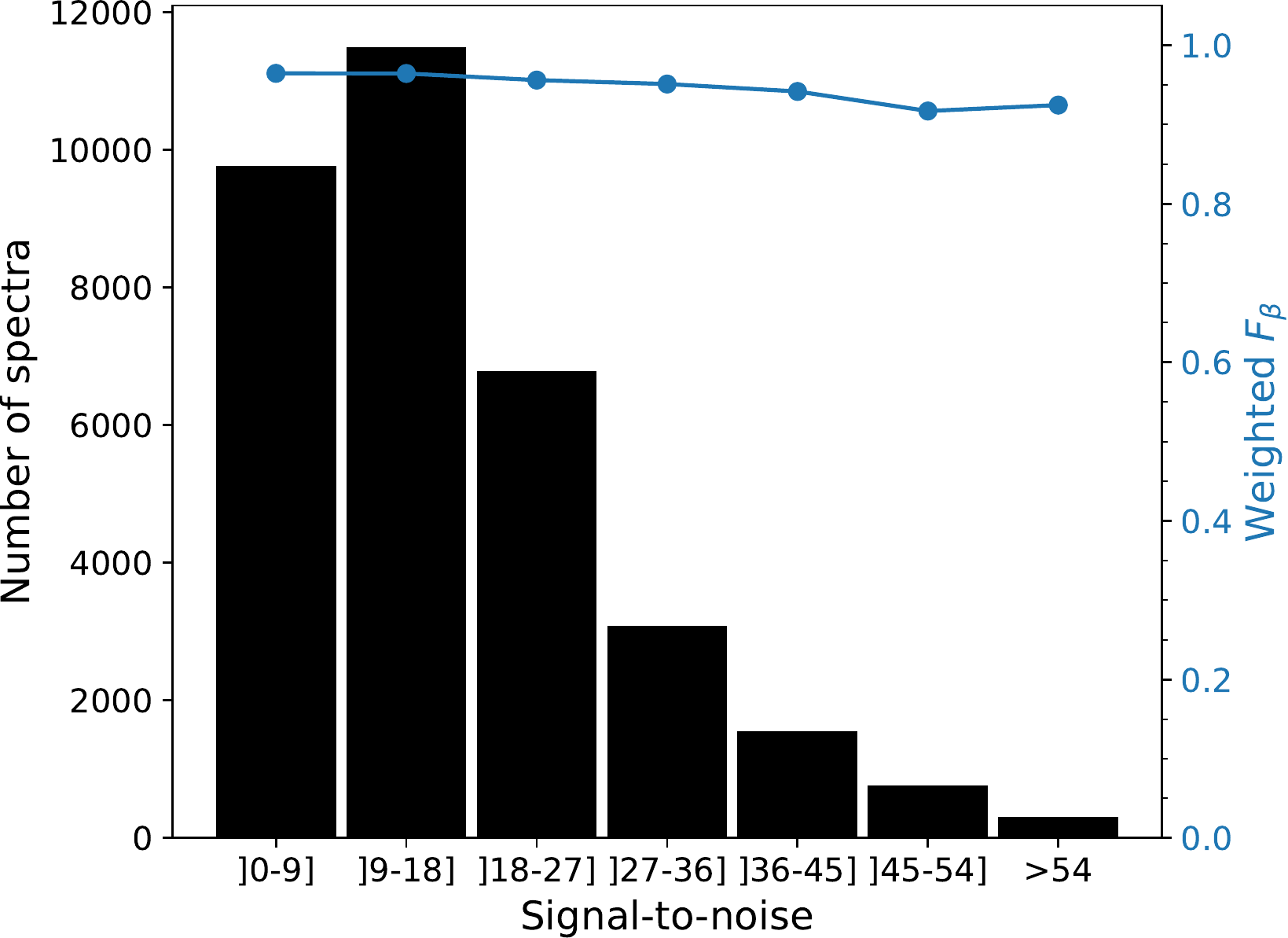}
    \includegraphics[width=\columnwidth]{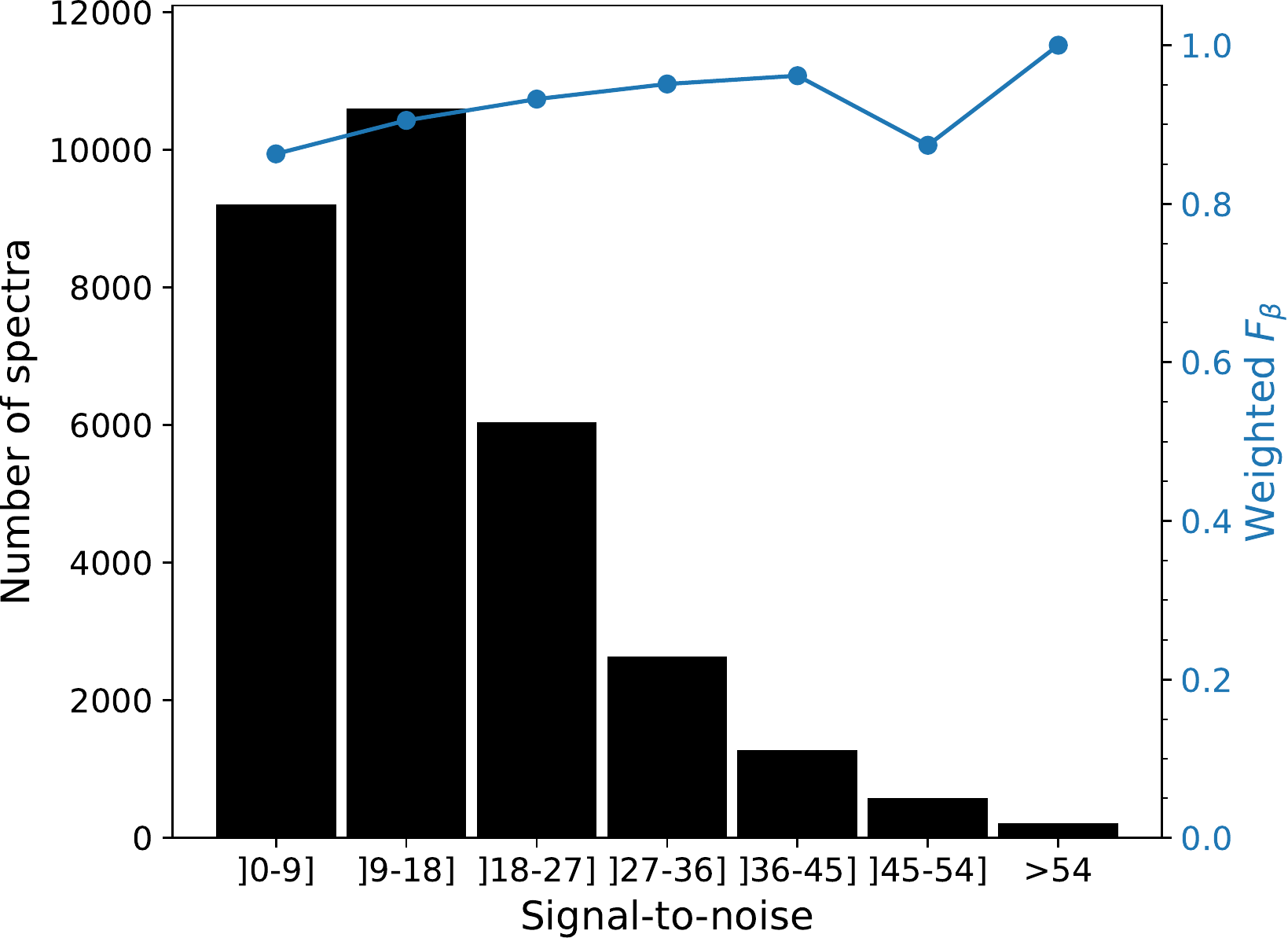}
    \caption{SNR histograms of spectra used to train and to test the spectroscopic classification modules of our pipeline, overlaid with the class-weighted average $\Fbeta$-score. The top panel is for the primary spectroscopic type classification module, while the bottom is for the main sequence companion module.}
    \label{fig:snrbins}
\end{figure}

To evaluate the performance of the primary spectroscopic type module in different SNR bins, we plot the SNR histogram of all spectra in the upper panel of Figure \ref{fig:snrbins}, including white dwarfs with secondary types and main sequence binaries discussed previously, and overlay the class-weighted average of the $\Fbeta$-score. The primary spectroscopic type module shows an excellent score at $\Fbeta\sim0.95$ for spectra under SNR $\sim$45, with a slight drop to $\sim$0.91 for higher SNR. Though it may seem counter-intuitive that higher SNR spectra are slightly more difficult to classify for the neural networks, one must keep in mind the small number of spectra relative to the bulk at SNR 9-18, along with the fact that classes with few known objects -- thus more difficult to classify -- tend to have a higher SNR.

We find similar results for the main sequence companion module, for which we plot the histogram and class-weighted $\Fbeta$-score in the lower panel of Figure \ref{fig:snrbins} using the sample of spectra described in Section \ref{sec:ms}. The module also shows an excellent score of $\Fbeta\gtrsim0.9$ over the entire SNR $>9$ range, generally increasing along with higher SNR. This trend differs from that found for the primary spectroscopic type module because the detection of MS companions becomes easier with better signal, but also because the classification is set as a binary problem (WD vs WD$+$MS), removing the effect that might have been caused by rare classes whose spectra are mostly found at high SNR.

\begin{table}
 \caption{Sample of low SNR spectra used to test the main spectroscopic type classification module}
 \label{tab:lowsnr}
 \centering
 \begin{tabular}{lrrrr}
    \hline
    Label & $N$ & $N_\mathrm{agree}$ & $N_\mathrm{uncert}$ & $N_\mathrm{disagree}$\\
    \hline
    DA & 7531 & 7438 & 45 & 48 \\
    DC & 1241   & 995 & 97 & 149 \\
    DB & 331   & 321 & 5 & 5 \\
    DZ & 362    & 315 & 20 & 27 \\
    DQ & 47    & 41 & 4 & 2 \\
    DAH & 15   & 10 & 0 & 5 \\
    DO & 1 & 0 & 0 & 1 \\
    hotDQ & 7  & 5 & 2 & 0 \\
    PG1159 & 0 & 0 & 0 & 0 \\
    sdB & 20   & 2 & 6 & 12 \\
    sdOB & 3  & 0 & 1 & 2 \\
    sdO & 8   & 0 & 4 & 4 \\
    CV & 75    & 72 & 1 & 2 \\
    \hline
    Total & 9734 & 9292 & 185 & 257\\
  \hline
 \end{tabular}
\end{table}

\begin{figure}
    \centering
    \includegraphics[width=\columnwidth]{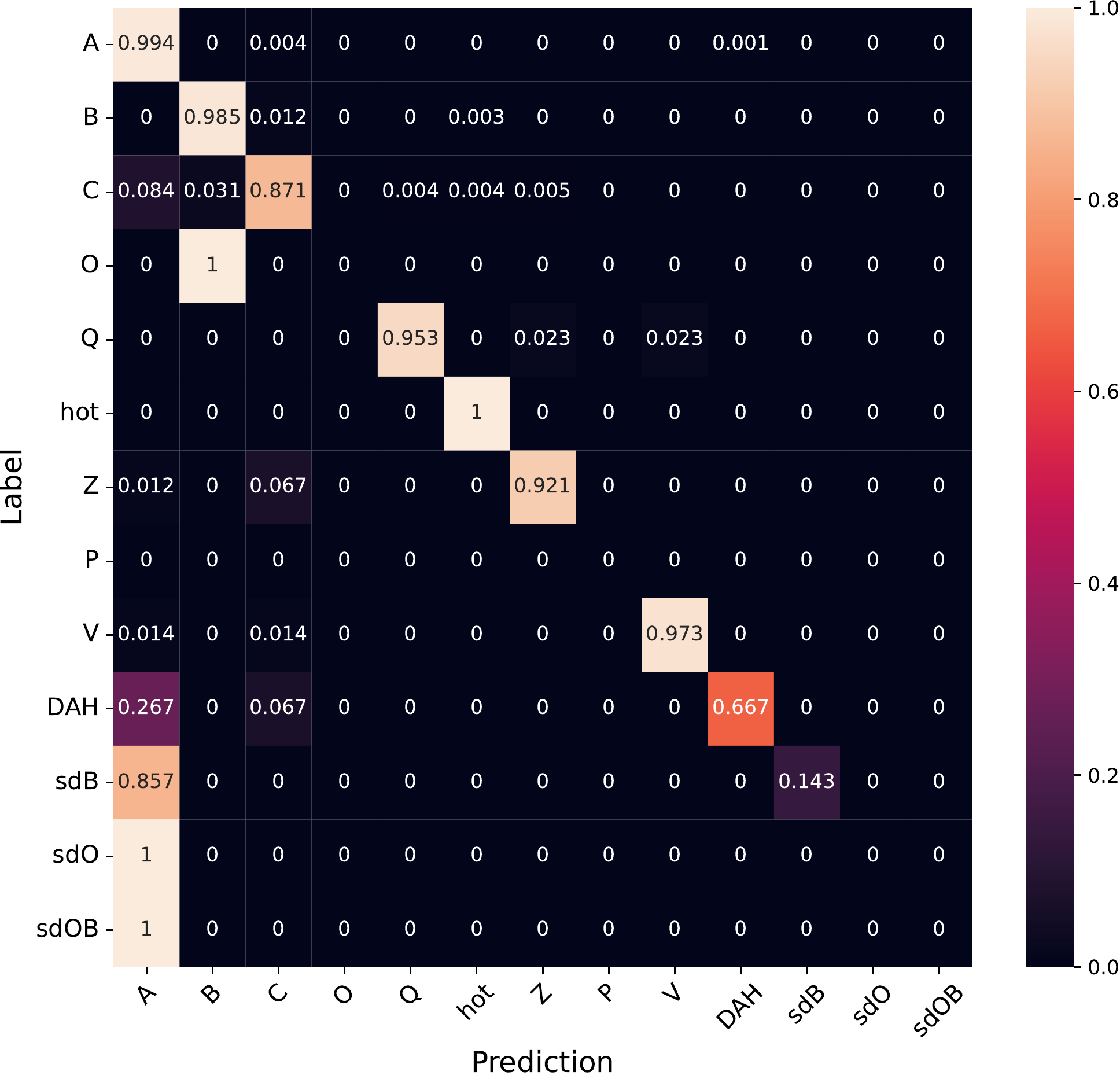}
    \caption{Primary spectroscopic type confusion matrix of confident predictions ($\Pclass>0.6$) for SNR$\leq9$ spectra.}
    \label{fig:cmatrixlowsnr}
\end{figure}

Even though we do not use low-SNR ($\leq 9$) spectra to train and to test the neural networks of our pipeline, they constitute $\sim$30\% of all spectra in the GF21 catalogue, warranting at the very least a short performance assessment of the spectroscopic classification modules in this SNR regime. We verify the global performance by including a bin containing SNR $\leq9$ spectra in the histograms displayed in Figure \ref{fig:snrbins}. Both modules display very high scores on par with higher SNR spectra, with the primary spectroscopic type module showing a $\sim$0.96 score, and the main sequence companion module showing a 0.86 score. We further study the high score of the primary spectroscopic type module with its confusion matrix in Figure \ref{fig:cmatrixlowsnr}, showing a $\gtrsim$90\% agreement for most classes when applying the $\Pclass>0.6$ threshold. The only exception is for subdwarfs, for which the ensemble predicts a DA type for 27 out of the 31 spectra. These classes are known to be notoriously difficult to distinguish in the low SNR regime. The confusion between DA and DAH also remains present, with 4 of the 15 labelled DAH being predicted as DA. A more detailed list of low-SNR spectra used to test the main spectroscopic type module and their resulting classification is provided in Table \ref{tab:lowsnr}.

We conclude that the spectral classification modules of our pipeline are reliable over the entire SNR range, including noisier spectra (SNR$<9$) on which the networks were not trained. We caution, however, that neural networks are known for producing overconfident predictions on out-of-distribution data \citep{nguyen2014, goodfellow2014}, and recommend that extra care be taken when interpreting results in low-SNR regimes.

\section{Gaia-SDSS White Dwarf Catalogue}\label{sec:dr17}

In this section, we use our pipeline to identify white dwarf candidates from a large sample of Gaia objects and classify them spectroscopically using SDSS DR17 spectra. The results are made available as an online catalogue, along with recommendations on how to use it, as well as the pipeline itself.

\subsection{Candidate selection}\label{sec:catcnd}

\begin{figure}
    \centering
    \includegraphics[width=\columnwidth]{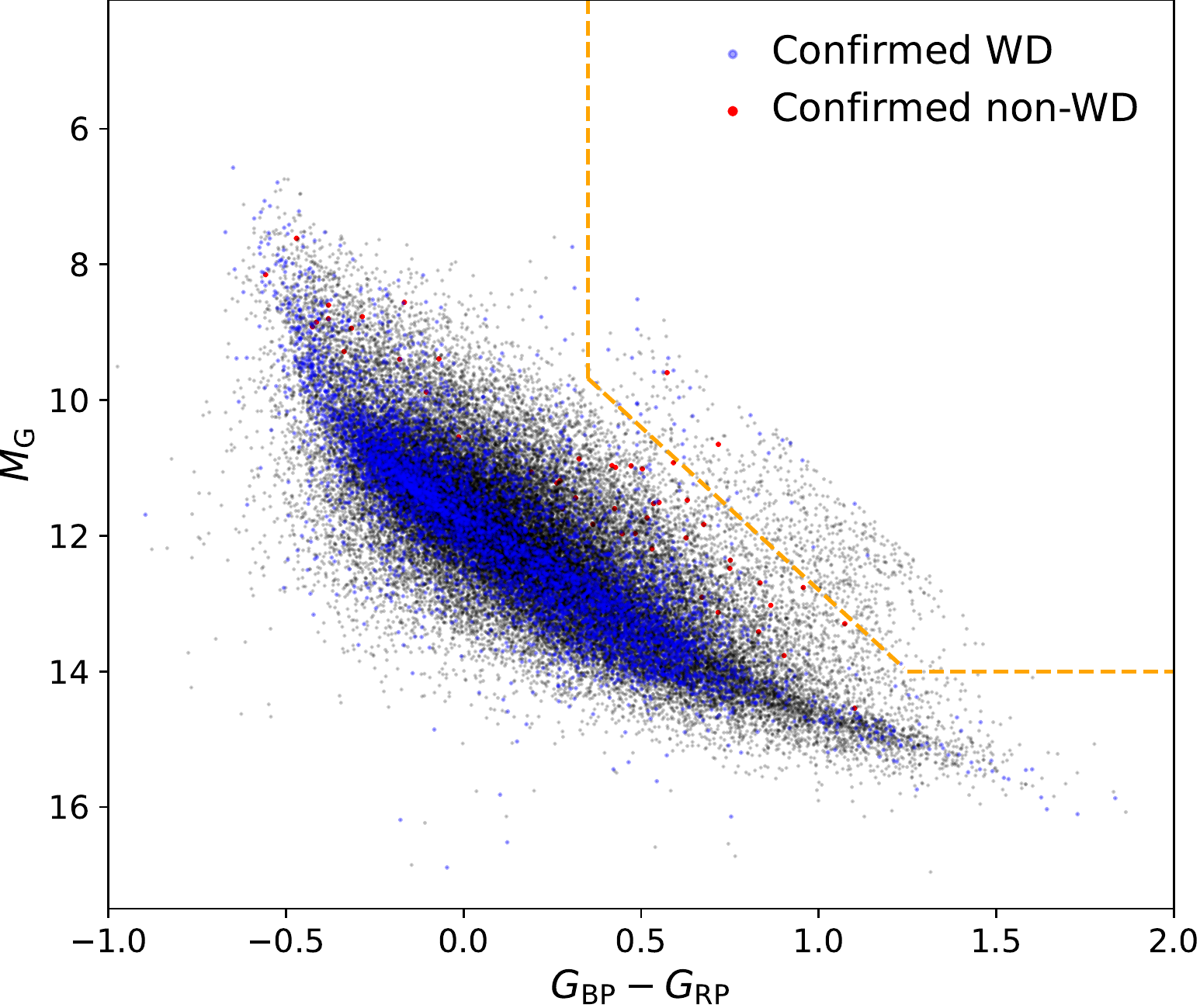}
    \caption{Gaia HRD of white dwarf candidates selected among 1.3M Gaia objects. Candidates are shown in grey, while spectroscopically confirmed white dwarfs and non-white dwarfs are shown in blue and red, respectively. Objects found to the right of the region delimited by the orange dashed lines are likely WD$+$MS objects (see Section \ref{sec:catcnd}).}
    \label{fig:hrdcnd}
\end{figure}

% Candidate selection
As our starting point, we calculate the probability of being a white dwarf for the 1.3M objects found in the GF21 Gaia main catalogue. We apply a probability threshold of $\Pwd>0.75$ and an uncertainty limit of 0.02, resulting in 424 096 white dwarf candidates. Of these candidates, 25 205 are spectroscopically confirmed white dwarfs, and 50 are non-WD according to the MWDD and/or the GF21 SDSS catalogue. We reclassify as candidates 131 non-WD objects found within the white dwarf locus in the Gaia HRD since their spectra have very low SNR ($\lesssim5$) and are too noisy for reliable visual classification. We show in Figure \ref{fig:hrdcnd} the Gaia HRD of the candidates as well as spectroscopically confirmed white dwarfs and non-WD objects. For visibility purposes, a random selection of 15\% of the candidates and a quarter of the confirmed white dwarfs is displayed, while all non-WD objects are kept.

A large number of white dwarf candidates can be seen above the faint end of the white dwarf sequence, as delimited by the orange dashed line in Figure \ref{fig:hrdcnd}. This region is typically populated by WD$+$MS binaries \citep{rebassa2016,rebassa2021} and here we look for clues to confirm whether this is the case. We find that 2583 out of the 4311 candidates above the line have renormalized unit weight error (\texttt{ruwe}) greater than 1.1, a quantity representing the quality of the astrometric solutions, for which a value between $1.1\lesssim\texttt{ruwe}\lesssim1.4$ has been found to indicate possible movement perturbations caused by an unresolved companion \citep{belokurov2020}. An additional 583 candidates show red flux excess ($\texttt{phot\_bp\_rp\_excess\_factor}\gtrsim1.3$) that is at least 0.2 larger than those of other objects with a similar absolute $G$ magnitude. \citet{riello2021} found that many objects with large excess factors tend to either have emission lines in the wavelength range where the RP passband has a larger transmissivity with respect to the $G$ passband, or to be blended sources. High values of \texttt{ruwe} and \texttt{phot\_bp\_rp\_excess\_factor} may thus imply that the white dwarf candidates inside the region delimited by orange dashed lines in Figure \ref{fig:hrdcnd} are genuine white dwarf binary systems. For ease of selection (or removal), we include a binary flag \texttt{above\_locus} in our catalogue for the 4317 objects located within the region.

% compare with gf21
We compare our selection of candidates to the GF21 main catalogue by extracting all objects with $P_\mathrm{WD, GF21}>0.75$. We find our selection to contain 25 816 candidates not in the GF21 selection, including 869 spectroscopically confirmed white dwarfs and 28 non-WD objects, the former set mainly consisting of DA and DA$+$MS white dwarfs. There are 48 candidates present in the GF21 selection, but not in ours, including 28 confirmed non-WD and no confirmed white dwarfs.

% compare with gcns
We also compare our candidate selection within 100 parsecs of the Sun to the Gaia Catalogue of Nearby Stars \citep[GCNS,][]{smart2021}, which provides a probability for objects to be white dwarfs using a Random Forest algorithm. Apart from the choice of algorithm, the main difference between our two methods lies in the choice of training data. Our training set is restricted to the white dwarf region in the Gaia HRD through color cuts, while \citet{smart2021} use the entire HRD space. Moreover, our non-WD training examples are solely spectroscopically confirmed objects, whereas the GCNS uses any object that is not a spectroscopically confirmed white dwarf as part of their non-WD training set, which may include yet-to-be confirmed white dwarfs. We compare objects in the GCNS with a probability of being a white dwarf above 0.75 and find a single candidate not present in our own selection, but we find 366 candidates present in our selection but missing in the GCNS, among which 107 are spectroscopically confirmed white dwarfs and no confirmed non-WD objects. Overall, our candidate selection appears to be more complete while also being less contaminated than other catalogues.

\subsection{Spectroscopic classification}

% Spectroscopic sample
Optical spectra for the white dwarf candidates are obtained by calculating the position of the Gaia objects at the J2000 epoch using their proper motions, then by cross-matching them with the SDSS DR17  the latest and last data release of SDSS-IV \citep{blanton2017}. This data release includes new observations through January 2021, as well as updates to some calibration files affecting all eBOSS spectra taken after the summer of 2017. Therefore, all spectra after MJD 58000 are different from their equivalent DR16 version and can be considered as unseen data for the networks. We also supplement our sample with 3591 spectra from the MWDD and the GF21 SDSS catalogue missed by our cross-match procedure. We remove all spectra that do not have the full coverage of 3842-7000 \AA\ required by our networks, leaving a total of 36 523 spectra belonging to 27 866 unique Gaia white dwarf candidates. 

We pass all the spectra through the primary spectroscopic type classification module and present the results in Table \ref{tab:NNmain}, where $N^\mathrm{conf}_\mathrm{Gaia}$ is the number of unique Gaia objects for which their highest SNR spectrum has a confident prediction above the $\Pclass>0.6$ threshold, $N^\mathrm{uncert}_\mathrm{Gaia}$ is the number of unique Gaia objects whose predictions for their highest SNR spectrum fall below or equal to the threshold. Also included are the number of spectra predicted to belong to each class, $N^\mathrm{conf}_\mathrm{spec}$ and $N^\mathrm{uncert}_\mathrm{spec}$, following the same notation as for the Gaia columns. About 97.5\% of both spectra and unique Gaia objects were assigned a high probability primary spectral type, reducing the number of spectra requiring visual inspection from 36 523 to 872 in less than a minute.

A surprising result from the automated classification is the identification of 131 hotDQ Gaia objects, a much larger number than what is currently known in the literature \citep{dufour2008, koester2019, fusillo2021}. To our knowledge, 66 of these have already been identified as hotDQ by at least one study, 49 have previously been classified as another type than hotDQ (usually DAH or DQ), and the remaining 16 appear to be new discoveries. The large number of hotDQ may also be due to differing definitions of this class. Indeed, GF21 do not state what features were used to distinguish hotDQ from other DQ white dwarfs, and seem to include warmDQ \citep{dufour2013} in the hotDQ class. In their analysis of carbon-atmosphere white dwarfs, \citet{coutu2019} classify white dwarfs showing molecular carbon bands as DQ, neutral atomic carbon lines as warmDQ, and ionized carbon lines as hotDQ. Given these definitions, visual inspection of the spectra labelled and predicted as hotDQ reveals some of the objects may actually be warmDQ. A more detailed analysis of these objects would allow to confirm their true classes.

\begin{table}
 \caption{Primary spectral type classification predicted by our pipeline for white dwarf candidates in the Gaia-SDSS DR17 sample.}
 \label{tab:NNmain}
 \centering
\begin{tabular}{lrrrr}
\hline
Class &  $N^\mathrm{conf}_\mathrm{Gaia}$ &  $N^\mathrm{uncert}_\mathrm{Gaia}$ &  $N^\mathrm{conf}_\mathrm{spec}$ &  $N^\mathrm{uncert}_\mathrm{spec}$ \\
\hline
DA     &                            21434 &                                178 &                            28093 &                                211 \\
DB     &                             1887 &                                 35 &                             2758 &                                 43 \\
DC     &                             2007 &                                341 &                             2518 &                                416 \\
DO     &                               68 &                                  5 &                              102 &                                  5 \\
DQ     &                              265 &                                 21 &                              362 &                                 26 \\
hotDQ  &                              131 &                                 22 &                              181 &                                 26 \\
DZ     &                              896 &                                 80 &                             1120 &                                 96 \\
PG1159 &                               12 &                                  0 &                               17 &                                  1 \\
CV     &                              157 &                                  9 &                              225 &                                 12 \\
DAH    &                              174 &                                 17 &                              237 &                                 21 \\
sdB    &                               13 &                                  5 &                               18 &                                  6 \\
sdO    &                                6 &                                  0 &                                9 &                                  1 \\
sdOB   &                                9 &                                  7 &                               11 &                                  8 \\
\hline
\end{tabular}
\end{table}

% +MS classification
As a final step, all candidates classified as DA, DB, and DC white dwarfs by our neural networks are passed through the third classification module to determine whether they have a MS companion. Assuming a probability threshold of 0.5, we find 1380 spectra showing signs of companionship, a number comparable to those identified by GF21. To verify how the pipeline deals with WD$+$MS systems, we compare our results with the spectroscopic catalogue of WD$+$MS binaries in SDSS DR12 published by \cite{rebassa2016}. We first cross-match the 979 SDSS objects in their catalogue with the Gaia DR3 and find 258 within the 1.3M objects in the GF21 main catalogue. The list of white dwarf candidates produced by our candidate selection module contains 211 of these, all of which were correctly classified as WD$+$MS by our pipeline. 

Binary systems were excluded from the training sets of both the candidate selection and the main spectroscopic type modules. Consequently, our pipeline may not be the optimal tool for discovering such objects. However, the MS companion module could be used as a stand-alone classifier to identify new WD$+$MS binaries in samples with appropriate selection criteria. The module could also benefit from training on synthetic data, as current WD$+$MS samples are strongly biased towards relatively equal contribution from both members \citep{rebassa2021}, and may prove to be able to classify objects for which visual inspection is too ambiguous.

\section{Data Availability}\label{sec:maincat}
The results of Section \ref{sec:dr17} are available on the MWDD website\footnote{montrealwhitedwarfdatabase.org} and in the VizieR catalogue access tool as two catalogues. The first catalogue includes the probability of being a white dwarf for the 1.3M Gaia objects in Section \ref{sec:maincat}, along with all Gaia parameters used for our analysis and discussion. See Table \ref{tab:catcnd} for a list of column names and description. The second catalogue contains the list of objects for which SDSS spectra were found, as well as their spectroscopic classification results, in addition to the same columns as in the Gaia candidate catalogue. We provide the \texttt{fiberid}, \texttt{mjd} and \texttt{plateid} as a way to identify the spectra in the SDSS database. Note that Gaia objects may have multiple spectra, or vice versa. A list of the columns unique to the results of the Gaia-SDSS catalogue is shown in Table \ref{tab:catsdss}. 

\begin{table*}
 \caption{Columns unique to our Gaia candidate catalogue}
 \label{tab:catcnd}
 \centering
 \begin{tabular}{lll}
    \hline
    Column Header & Description \\
    \hline
    \texttt{source\_id}  & Unique source identifier in Gaia DR3 \\
    \texttt{P\_wd}  & Probability of being a white dwarf (see Section \ref{sec:meth})\\
    \texttt{P\_wd\_u}  & Uncertainty on the probability of being a white dwarf\\
    \texttt{above\_locus}  & Binary flag indicating whether the object is above the white dwarf locus and may be a WD$+$MS binary (see Section \ref{sec:catcnd})\\
  \hline
 \end{tabular}
\end{table*}
\begin{table*}
 \caption{Columns unique to the Gaia-SDSS DR17 catalogue. All columns in the Gaia candidate catalogue are also included, but not shown here.}
 \label{tab:catsdss}
 \centering
 \begin{tabular}{lll}
    \hline
    Column Header & Description \\
    \hline
    \texttt{P\_\{class\}}  & Probability of the spectrum being of primary spectroscopic type \{class\} (see Section \ref{sec:meth} for the 13 possible classes)\\
    \texttt{P\_\{class\}\_u}  & Uncertainty on the probability of the spectrum being of primary spectroscopic type \{class\}\\
    \texttt{P\_ms}  & Probability of the presence of a MS companion in the spectrum (see Section \ref{sec:meth})\\
    \texttt{P\_ms\_u}  & Uncertainty on the probability of the presence of a MS companion in the spectrum\\
  \hline
 \end{tabular}
\end{table*}

Granular control over the completeness and contamination rate for both the white dwarf candidate selection and the spectroscopic classifications can be achieved by imposing different prediction probability thresholds or prediction uncertainty limits. In principle, the prediction probability and uncertainty could be combined to look for spectra that show secondary spectroscopic signatures, though this has not been tested here. For a good balance between completeness and contamination, we recommend the following thresholds: $\Pwd>0.75$ and an uncertainty limit of 0.02 for candidate selection, $\Pclass>0.6$ for the primary spectroscopic type with the highest prediction probability, and $\Pms>0.5$ for MS companion detection. We also remind the reader to be cautious when interpreting predictions for spectra with low SNR ($\lesssim9$), as the neural networks may be overconfident in their predictions (see Section \ref{sec:res}).

The modules of the pipeline are made available to try and to use on the MWDD website\footnote{montrealwhitedwarfdatabase.org/MLTools}. A description of the required inputs, file formats can be found there. Future versions of the pipeline and machine learning tools will also be uploaded on this page. By making these publicly available, we aim to facilitate the transition from the traditionally manual approaches of white dwarf analysis to rapid, automated statistical tools for the entire community, and encourage other teams to do the same. As the field of astronomy enters a new era of Big Data, collaborative efforts will be more important than ever to ensure science is not hampered by sheer quantity of observations, and to allow astronomers to focus on interesting cases waiting to be discovered.

\section{Conclusion}\label{sec:conc}
In this paper, we presented a fully automated, data-driven pipeline for white dwarf candidate selection and spectroscopic classification based on neural network ensembles. The pipeline is composed of three modules that can be used independently for a variety of purposes. The first module calculates the probability of being a white dwarf given Gaia photometric and astrometric data, and correctly identified $>99\%$ of white dwarfs in the test set, with a contamination rate of $\sim$1.4\%. The second module predicts the primary spectroscopic type of a white dwarf given an optical spectrum with $>90\%$ precision for most classes according to cross-validation tests. The last module calculates the probability of main sequence star contamination being present in the spectrum with $>91\%$ precision on its test set. The two spectroscopic modules were trained with SDSS DR 16 spectra.

We applied our pipeline to 1.3M Gaia objects located in, or near the white dwarf locus in the Gaia HRD and found 424 096 high probability white dwarf candidates for which we cross-matched 36 523 SDSS DR17 spectra, creating the first white dwarf catalogue with quantifiable spectroscopic classifications. The entire process is orders of magnitude faster than the current manual inspection approach, taking about 10 minutes on a Mac M1 laptop, with about 9 minutes taken by the candidate selection module. In addition to the benefits of quantifiable classifications and speed, neural networks remove the need to select manually the relevant features by learning which ones best distinguish one class from the others.

The pipeline presented here will be particularly useful for the SDSS-V, as the spectroscopic classification modules are already trained on data observed by the same instruments. The pipeline can also serve as a base model for other surveys (e.g., DESI, 4MOST, LAMOST), where fine-tuning and transfer learning methods can be applied to adapt the spectroscopic modules to the new data distribution. Deep ensembles seem to offer benefits in out-of-distribution settings \citep{ovadia2019,gustafsson2020}, although with some limitations \citep{rahaman2021}, and may require only small adjustments to perform well on data from surveys other than SDSS. Future work will aim to provide secondary spectroscopic feature identification and improved classification for rare classes such as PG1159 and hotDQ. In particular, the addition of synthetic spectra to augment the training data offers a promising avenue that will be investigated. Such machine learning tools will soon become indispensable as observations from the next generation of spectroscopic surveys will provide millions of spectra for hundreds of thousands of white dwarf candidates.

\section*{Acknowledgements}
%money
This work was supported in part by the NSERC Canada and by the Fund FRQ-NT (Québec).
%gaia
This work has made use of data from the European Space Agency (ESA) mission {\it Gaia} (\url{https://www.cosmos.esa.int/gaia}), processed by the {\it Gaia} Data Processing and Analysis Consortium (DPAC, \url{https://www.cosmos.esa.int/web/gaia/dpac/consortium}). Funding for the DPAC has been provided by national institutions, in particular the institutions participating in the {\it Gaia} Multilateral Agreement.
%SDSS
Funding for the Sloan Digital Sky Survey (\url{https://www.sdss.org}) has been provided by the Alfred P. Sloan Foundation, the U.S. Department of Energy Office of Science, and the Participating Institutions. SDSS-IV acknowledges support and resources from the Center for High-Performance Computing at the University of Utah, and is managed by the Astrophysical Research Consortium for the Participating Institutions of the SDSS Collaboration.
% databases
This research has also made use of the NASA Astrophysics Data System Bibliographic Services; the Montreal White Dwarf Database (Dufour et al. 2017); the SIMBAD database, operated at the Centre de Données astronomiques de Strasbourg (Wenger et al. 2000)

\section*{Data Availability}
A full description of the data and software availability is provided in Section \ref{sec:maincat}.


\bibliographystyle{mnras}
\bibliography{ms.bib}

%%%%%%%%%%%%%%%%% APPENDICES %%%%%%%%%%%%%%%%%%%%%

\appendix

\section{Neural network architectures}\label{app:arch}
The architecture details for our three pipeline modules are briefly described here. All modules are implemented using the \texttt{tensorflow} (version 2.10) Python library \citep{tensorflow2015}. For every network, regardless of the module, we use an initial learning rate of 0.01 and the \texttt{ReduceLROnPlateau} callback with a \texttt{factor} of 0.33, \texttt{min\_delta} of $10^{-3}$, and \texttt{patience} of 3. We use the \texttt{tensorflow} implementation of the Adam optimizer \citep{kingma2014}.

Starting with the candidate selection module, each network is composed of 5 fully-connected layers with 56 hidden units in the first layer, doubling up for each successive layer, followed by a single-unit fully-connected layer that outputs the probability of an object being a white dwarf. The output layer has a \texttt{sigmoid} activation function, while all other layers use a \texttt{LeakyReLU} activation \citep{Maas2013} with the leak parameter set to 0.1, along with \texttt{dropout} \citep{Srivastava2014} set to 0.5. We train the networks for 100 epochs using a binary cross-entropy loss.

The primary spectroscopic type module neural networks use a mix of fully-connected and convolutional \citep{Krizhevsky2012} layers. Each network has 4 convolutional layers with a kernel size of 5, 14 feature maps and a stride of 2, followed by a fully-connected layer with 100 hidden units, and a final fully-connected layer with 13 hidden units for the output. The output layer has a \texttt{sigmoid} activation function, while all other layers use a \texttt{LeakyReLU} activation with the leak parameter set to 0.1, along with \texttt{dropout} set to 0.4. We also apply \texttt{dropout} to the input with a 0.2 probability. Each network is trained for 30 epochs using a categorical cross-entropy loss.

The main sequence companion detection module neural networks have 3 fully-connected layers with 512 hidden units in the first layer, doubling down for every successive layer, followed by a single-unit fully-connected output layer. The output layer has a \texttt{sigmoid} activation function, while all other layers use a \texttt{LeakyReLU} activation with the leak parameter set to 0.1, a \texttt{dropout} set to 0.5, and L1L2 kernel regularization with default parameters. We also apply \texttt{dropout} to the input with a 0.2 probability. We train the networks for 100 epochs using a binary cross-entropy loss.

%%%%%%%%%%%%%%%%%%%%%%%%%%%%%%%%%%%%%%%%%%%%%%%%%%

% Don't change these lines
\bsp	% typesetting comment
\label{lastpage}
\end{document}